\documentclass[review]{elsarticle}

\usepackage{lineno,hyperref}
\usepackage{amsmath}
\usepackage{caption}
\usepackage{subcaption}
\modulolinenumbers[5]

\journal{Superlattices and Microstructures}

\begin{document}

\begin{frontmatter}

\title{Andreev bound states in Superconductor-Quantum dot Josephson junction at infinite-U limit}

\author{Tanuj Chamoli \fnref{myfootnote1}}
\author{Ajay \fnref{myfootnote2}}
\address{Department of Physics, Indian Institute of Technology Roorkee : 247667, Uttarakhand, India}
\fntext[myfootnote1]{tchamoli@ph.iitr.ac.in}
\fntext[myfootnote2]{ajay@ph.iitr.ac.in}


\begin{abstract}
Andreev bound states (ABSs) are studied in quantum dot coupled to conventional BCS superconducting leads on the basis of effective slave boson Hamiltonian in an infinite-U (Coulomb interaction) limit followed by Green's function technique. From the relevant Green's function, density of states (DOS) is analyzed  at different superconducting phase differences. On the basis of numerical computation, it is pointed out from DOS plot that ABSs (sub-gap states)  arise due to the phase difference between left and right superconducting leads. Due to the dependence on superconducting phase difference, ABSs are current carrying states. We have also analyzed the energy of ABSs as a function of phase difference between left and right superconducting lead for electron hole symmetric as well as non symmetric cases at various dot-lead coupling strengths. It is also pointed out that a finite internal gap arises between upper and lower Andreev bound states in the absence of electron-hole or left-right symmetry. These results are viewed in the light of  existing theoretical analysis.
\end{abstract}

\begin{keyword}
\texttt \sep Quantum dot\sep BCS Superconductor\sep Andreev bound states\sep Infinite-U slave Boson mean field approach
\MSC[2020] 81Q37\sep  81T99 \sep 81V65 \sep 82D55
\end{keyword}

\end{frontmatter}

\section{Introduction}

Josephson transport across quantum dot junction has been widely used in nanoelectronic  devices  such as in superconducting quantum interference devices (SQUIDS) \cite{chui},  Cooper pair splitters ~\cite{Lh2} and superconducting quantum bits ~\cite{john} which are the basis for superconducting quantum computers. Josephson junction is obtained by sandwiching thin insulating layer between two superconducting leads and Josephson supercurrent flows across the junction due to the phase difference between Cooper pairs present on the left and right superconducting lead   ~\cite{Josephson}. Insulating layer can be replaced by some nanoscopic structures such as quantum dots (QDs) ~\cite{Reed}. QD itself, is not a superconducting but Cooper pairs tunneling can take place through it because coherence length of Cooper pairs is significantly larger as compared to the size of the QD. Due to the proximity effect ~\cite{Yashai}, induced pairing takes place on the dot around the Fermi level. In this way, a reduced gap is arised in the dot. Consequently, subgap states are observed in S-QD-S system which actually arises due to the superconducting phase difference between two leads ~\cite{yuzhu,Akumar}.  These states are called Andreev bound states and are responsible for supercurrent flowing through the junction. Andreev bound states are consequences of the phenomena called multiple Andreev reflection ~\cite{JDPillet}. 

For odd number of electrons occupation , QD acts as a magnetic impurity, and spin of quantum dot gets coupled to spin of one electron of the Cooper pair from the leads with a characteristic binding energy $k_BT_k$, where $T_K$ is the Kondo temperature. Kondo temperature for normal metallic lead varies as $ T_k \sim e^{-1/{\rho J}}$, where $J$ is the exchange interaction between spin of QD and spin of  one of the electron of metallic lead ~\cite{Kondo}. It leads to the formation of Kondo singlet in N-QD-N junction and it can be visualized as a peak in density of states.  Although for superconducting leads, there is an attaractive interaction between two electrons to form a bound Cooper pair. Therefore, interplay arises between multiple Andreev reflection and Kondo effect in S-QD-S junction ~\cite{Bumkim,Joshua,KJ}. It is observed experimentally that for $T_k >\Delta$, where $\Delta$ is binding energy of Cooper pair, there is always likelihood  of formation of ABSs in the dot. It give rise to sharp increase in DOS within superconducting gap at $+\omega_s$ and $-\omega_s$, where $\omega_s$ corresponds to energy of Andreev bound states. Therefore, the ratio $ \Delta/T_k$ is an important parameter for the transmission of Cooper pair from left to right lead ~\cite{Rok}.  For even number of electrons occupation on dot, positive supercurrent while for odd number of electrons occupation negative supercurrent flows through the junction. It indicates that the supercurrent reverses the sign by adding an electron on the QD ~\cite{Jorden}.  

\begin{figure*}[ht]
\centering
\includegraphics[trim = 0.5cm 0.2cm 0.5cm 0.2cm, clip, 
width=0.5\textwidth]{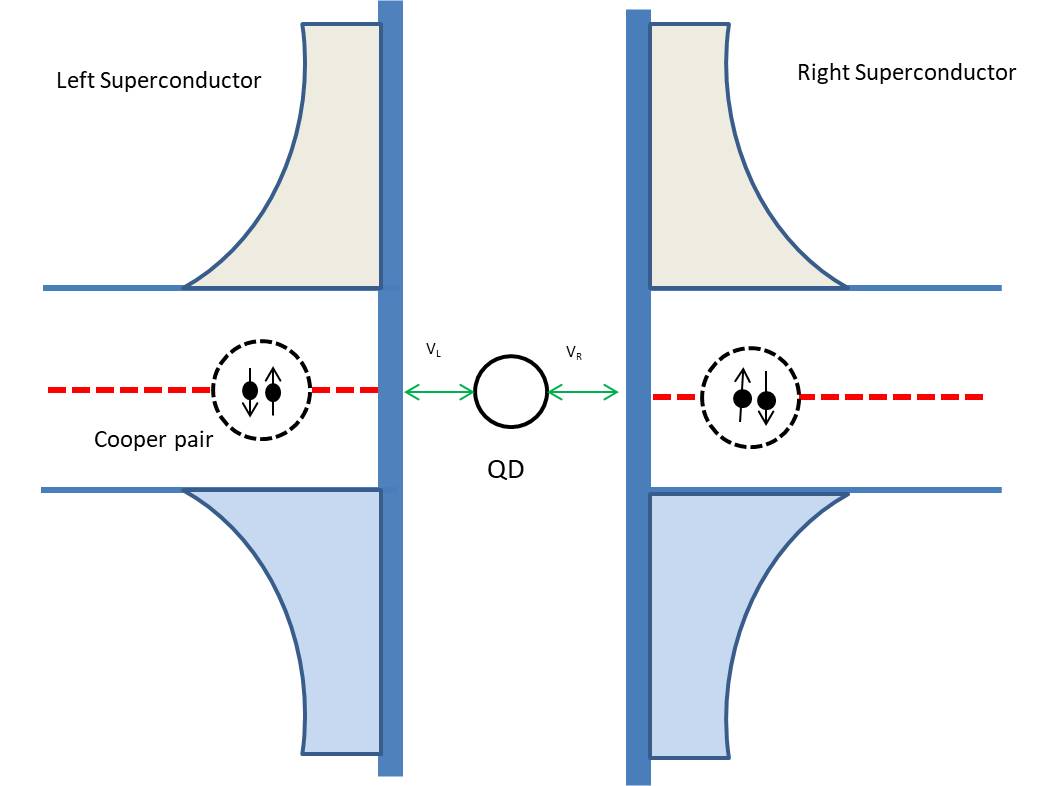}
\caption{Schematic of S-QD-S junction, showing coupling of QD with conventional superconducting leads with respect to Fermi level (shown as red dashed line)}
\label{schematic2} 
\end{figure*}

 As of now, several experimental  attempts  have been made to fabricate and characterize nanowire devices in which nanostructure is coupled to superconducting leads  ~\cite{silvano,H,AYu,HI}. These experiments indicate the existence of subgap states in S-QD-S system at various parameter regime.Further it is analyzed that these subgap states are effected by superconducting phase difference and dot-lead coupling strength. Outcomes of these experiments are analyzed by various analytical and computational techniques, which are based on different approximations. To mention few of these theoretical techniques include renormalization group approach ~\cite{Pillet,Minchu,Tanaka}, quantum monte carlo simulation ~\cite{DJ,RD},  non-crossing approximation ~\cite{Aashish,Martin,satoshi}, perturbation theory ~\cite{Evecino} and also non-equlibrium Green's function ~\cite{JF,Gagan}. Although infinite-U slave boson mean field (SBMF) method to understand the influence of infinite Coulomb interaction on the QD state connected to BCS superconducting leads is addressed by few of the researchers ~\cite{YAvishai,ALYeyati}, but as of now , role of strong onsite interaction on the formation of subgap states and on nature of Josephson transport as well is not clearly understood from theoretical point of view. Also extension of S-QD-S junction to multiple coupled QDs is yet to be analyzed. Therefore, in present attempt we have planned to address the issue related to strong on site Coulomb interaction (infinite-U) on the ABSs and their phase dependence, where we have used Bogolibouv transformation followed by infinite-U slave Boson mean field approach to understand the nature of S-QD-S junction at infinite Coulomb interaction. In the preceding section, we have presented the theoretical model for S-QD-S system.

\section{Theoretical formulation}

As depicted in Fig. \ref{schematic2} the model Hamiltonian for S-QD-S  junction where QD has single level is modelled by single impurity Anderson Hamiltonian (SIAM). BCS term is introduced to take care of superconducting leads. 

\begin{equation}
      H=H_{leads} + H_{QD} +H_{dot-lead} 
\end{equation}
where contribution of leads is given by
  \begin{equation}
   H_{leads} = \sum_{\alpha k\sigma}{\epsilon_{\alpha k\sigma} {c^\dagger}_{\alpha k\sigma} c_{\alpha k\sigma}}  -\sum_{\alpha k\sigma} ({\Delta_{k\alpha}   {c^\dagger}_{\alpha k\sigma} {c^\dagger}_{\alpha -k-\sigma} +h.c. )}  
   \end{equation}
   where $\epsilon_{k\alpha \sigma}$ denotes the energy of an electron with spin $\sigma$ ($\sigma=\uparrow, \downarrow$ denotes spin of spin up and spin down electrons respectively) in the $k^{\text{th}}$  energy state of lead  $\alpha$ ($\alpha=L,R$ corresponds to left and right lead). $\Delta_{k\alpha}=\left| \Delta_{k\alpha} \right|e^{i \phi_{k\alpha}}$ is complex superconducting order parameter in which $\phi_{k\alpha}$ is phase correponding to lead $\alpha$. For BCS, S-wave superconductor, superconducting order parameter does not have k dependency. Quantum Dot contribution is given as follows.
    \begin{equation}
   H_{QD} = \sum_{\sigma}\epsilon_d {d^\dagger}_{\sigma} d_{\sigma}
   \end{equation}
   $\epsilon_d$ is the energy level of the quantum dot. Coupling between dot level and superconducting lead is of the form
  \begin{equation}
   H_{dot-lead} = \sum_{\alpha  k\sigma}(h_{k \alpha}{c^\dagger}_{\alpha k\sigma} d_{ \sigma}+h_{k \alpha}^* {d{^\dagger}}_{ \sigma} c_{\alpha k \sigma})
   \end{equation}
where $h_{k\alpha}$ is coupling strength between dot and lead. To diagonalize the Hamiltonian Bogoliubov transformation is used, defined as follows

\begin{equation}
     c_{k \alpha \uparrow}=u_{k \alpha}^* \beta_{{k \alpha} \uparrow}+v_{k \alpha} {\beta^\dagger}_{-{k \alpha} \downarrow}
 \end{equation}
 
 \begin{equation}
     c_{-k \alpha \downarrow}=u_{k \alpha}^* \beta_{-{k \alpha} \downarrow}-v_{k \alpha} {\beta^\dagger}_{{k \alpha} \uparrow}
 \end{equation}
 
  \begin{equation}
       {c^\dagger}_{k \alpha \uparrow}=u_{k \alpha} {\beta^\dagger}_{{k \alpha} \uparrow}+v_{k \alpha}^* {\beta}_{-{k \alpha} \downarrow}
 \end{equation}
 
 \begin{equation}
     {c^\dagger}_{-k \alpha \downarrow}=u_{k \alpha} {\beta^\dagger}_{-{k \alpha} \downarrow}-v_{k \alpha}^* {\beta}_{{k \alpha} \uparrow}
 \end{equation}
 
 where
\begin{equation}
    \begin{split}
     {|u_{k \alpha}|}^2=\frac{1}{2}\bigg(1+\frac{\epsilon_{k \alpha}}{\sqrt{{\epsilon_{k\alpha}}^2+{|\Delta_{k\alpha}|}^2}}\bigg)\\{|v_{k \alpha}|}^2=\frac{1}{2}\bigg( 1-\frac{\epsilon_{k \alpha}}{\sqrt{{\epsilon_{k \alpha}}^2+{|\Delta_{k \alpha}|}^2}}\bigg)\\
     u_{k \alpha}^* v_{k \alpha}=\frac{\Delta_{k \alpha}}{2E_{k \alpha}}\\
     v_{k \alpha}^* u_{k \alpha}=\frac{\Delta_{k \alpha}^*}{2E_{k \alpha}}
     \end{split}
\end{equation}

 After Bogoliubov transformation, Hamiltonian takes the form as follows.
 
 \begin{multline}
       H_{Bog} =  \sum_{\alpha k}{E_{\alpha k} ({\beta^\dagger}_{\alpha k\uparrow} \beta_{\alpha k\uparrow}}+{\beta^\dagger}_{\alpha -k\downarrow} \beta_{\alpha -k\downarrow})+\sum_{\sigma}\epsilon_d {d^\dagger}_{\sigma} d_{\sigma} \\ +  \sum_{\alpha  k} [h_{k \alpha} ( u_{k\alpha} {\beta^\dagger}_{k \alpha \uparrow}+v_{k \alpha}^* {\beta}_{-k \alpha \downarrow})d_\uparrow+(u_{k\alpha} {\beta^\dagger}_{k \alpha \downarrow}-v_{k\alpha}^* {\beta}_{-k\alpha \uparrow}) d_\downarrow] \\  +h_{k\alpha}^* [d^\dagger_\uparrow (u_{k\alpha}^* \beta_{k \alpha \uparrow}+v_{k \alpha} {\beta^\dagger}_{-k \alpha \downarrow})+d^\dagger_\downarrow (u_{k\alpha}^* \beta_{k \alpha\downarrow}-v_{k\alpha} {\beta^\dagger}_{-k \alpha \uparrow})]
\end{multline}
where
\begin{equation}
    E_{k \alpha}={\sqrt{{\epsilon_{k \alpha}}^2+{|\Delta_{k \alpha}|}^2}}
\end{equation}
Now to study the system at infiite-U employing slave boson mean field approach, auxiliary fermionic operators $ f_ \sigma$ and  $ {f^{\dagger}}_{ \sigma}$ are introduced in terms of auxiliary boson field operators $b$  and  $ {b^\dagger} $. 
 
 \begin{equation}
      {d_{ \sigma}} = {b^\dagger} f_{ \sigma}  
   \end{equation}
 
  \begin{equation}
      {d^\dagger}_{ \sigma} = {f^\dagger}_{ \sigma} b 
   \end{equation}  
   
  Due to infinite-U, single electron occupancy will be there on QD. Condition for single occupancy is given as
 
    \begin{equation}
            {b^\dagger} b + \sum_{\sigma} {f^\dagger}_{ \sigma} f_{ \sigma}=1
   \end{equation}
   
   This condition allows to apply the slave-boson technique. Hamiltonian now takes the form as
 
 \begin{multline}
   H^{'} = \sum_{\alpha k}{E_{\alpha k} ({\beta^\dagger}_{\alpha k\uparrow} \beta_{\alpha k\uparrow}}+{\beta^\dagger}_{\alpha -k\downarrow} \beta_{\alpha -k\downarrow})   +\sum_{\sigma}\epsilon_d {f^\dagger}_{\sigma} b b^ \dagger f_{\sigma} \\  
      +\sum_{\alpha   k} h_{k \alpha}[(u_k {\beta^\dagger}_{k \uparrow}+{\beta}_{-k \downarrow})f_\uparrow b^\dagger +{f^\dagger}_\uparrow (u_k \beta_{k \uparrow}+v_k {\beta^\dagger}_{-k \downarrow})b]\\ +h_{k \alpha}^* [(u_k {\beta^\dagger}_{k \downarrow}-v_k {\beta}_{-k \uparrow}) f_\downarrow b^\dagger  +{f^\dagger}_\downarrow(u_k \beta_{k \downarrow}-v_k {\beta^\dagger}_{-k \uparrow})b]  + \lambda \sum_{\sigma} ({f^\dagger}_\sigma  f_\sigma +b^\dagger b - 1)
   \end{multline}
where $\lambda$ is Lagrange multiplier. In mean field approximation we replace Boson operators by their expectation values, $\langle b^\dagger(t) \rangle=\langle b(t) \rangle=b$. Effective Slave Boson mean field Hamiltonian finally comes out to be as

\begin{multline}
   H_{SBMF} = \sum_{\alpha k}{E_{\alpha k} ({\beta^\dagger}_{\alpha k\uparrow} \beta_{\alpha k\uparrow}}+{\beta^\dagger}_{\alpha -k\downarrow} \beta_{\alpha -k\downarrow})   +\sum_{\sigma} \tilde \epsilon_d {f^\dagger}_{\sigma}  f_{\sigma} \\  
      +\sum_{\alpha   k} \tilde{h_{k \alpha}} [(u_k {\beta^\dagger}_{k \uparrow}+{\beta}_{-k \downarrow})f_\uparrow  +{f^\dagger}_\uparrow (u_k \beta_{k \uparrow}+v_k {\beta^\dagger}_{-k \downarrow})]\\ +\tilde{h_{k \alpha}^*}[(u_k {\beta^\dagger}_{k \downarrow}-v_k {\beta}_{-k \uparrow}) f_\downarrow   +{f^\dagger}_\downarrow(u_k \beta_{k \downarrow}-v_k {\beta^\dagger}_{-k \uparrow})]  + \lambda (b^2 - 1)
   \end{multline}
where $\tilde {\epsilon_d} (= \epsilon_d +\lambda)$ is renormalized QD level and $\tilde {h_{K \alpha}} (=b h_{k \alpha}) $ is renormalized  dot lead coupling strength. Now, employing Green's function equation of motion, set of coupled equations is obtained. 

{{\begin{equation} \label{eq:17}
 (\omega-\tilde \epsilon_d) \langle \langle f_ \uparrow, {f^ \dagger}_ \uparrow \rangle \rangle = 1+\sum_{k \alpha} \tilde{h_{k \alpha}}^* {u_{k \alpha}}^* \langle \langle \beta_ {k \alpha \uparrow}, {f^ \dagger}_ \uparrow \rangle \rangle +\sum_{k \alpha} \tilde{h_{k \alpha}}^* v_{k \alpha} \langle \langle \beta^\dagger_{-k \alpha \downarrow}, {f^ \dagger}_ \uparrow \rangle \rangle
\end{equation}}}

{{\begin{equation}
 (\omega-E_{k \alpha}) \langle \langle \beta_{k \alpha \uparrow}, {f^ \dagger}_ \uparrow \rangle \rangle =  \tilde{h_{k \alpha}}  u_{k \alpha} \langle \langle f_ { \uparrow}, {f^ \dagger}_ \uparrow \rangle \rangle +\tilde{ h_{k \alpha}}^* v_{k \alpha} \langle \langle f^\dagger_{ \downarrow}, {f^ \dagger}_ \uparrow \rangle \rangle
\end{equation}}}

{{\begin{equation}
 (\omega+E_{k \alpha}) \langle \langle \beta^\dagger_{-k \alpha \downarrow}, {f^ \dagger}_ \uparrow \rangle \rangle = \tilde{ h_{k \alpha}} {v_{k \alpha}}^* \langle \langle f_ { \uparrow}, {f^ \dagger}_ \uparrow \rangle \rangle -\tilde{ h_{k \alpha}}^* {u_{k \alpha}}^* \langle \langle f^\dagger_{ \downarrow}, {f^ \dagger}_ \uparrow \rangle \rangle
\end{equation}}}

{{\begin{equation}
 (\omega+\tilde{\epsilon_d}) \langle \langle f^\dagger_ \downarrow, {f^ \dagger}_ \uparrow \rangle \rangle = \sum_{k \alpha} \tilde{h_{k \alpha}} {v_{k \alpha}}^* \langle \langle \beta_ {-k \alpha \uparrow}, {f^ \dagger}_ \uparrow \rangle \rangle -\sum_{k \alpha} \tilde{h_{k \alpha}} u_{k \alpha} \langle \langle \beta^\dagger_{k \alpha \downarrow}, {f^ \dagger}_ \uparrow \rangle \rangle
\end{equation}}}

{{\begin{equation}
 (\omega-E_{k \alpha}) \langle \langle \beta_{-k \alpha \uparrow}, {f^ \dagger}_ \uparrow \rangle \rangle = \tilde{ h_{k \alpha}} u_{k \alpha} \langle \langle f_ { \uparrow}, {f^ \dagger}_ \uparrow \rangle \rangle + \tilde{h_{k \alpha}}^* v_{k \alpha} \langle \langle f^\dagger_{ \downarrow}, {f^ \dagger}_ \uparrow \rangle \rangle
\end{equation}}}

{{\begin{equation} \label{eq:22}
 (\omega+E_{k \alpha}) \langle \langle \beta^\dagger_{k \alpha \downarrow}, {f^ \dagger}_ \uparrow \rangle \rangle = \tilde{ h_{k \alpha}} {v_{k \alpha}}^* \langle \langle f_ { \uparrow}, {f^ \dagger}_ \uparrow \rangle \rangle - \tilde{h_{k \alpha}}^* {u_{k \alpha}}^* \langle \langle f^\dagger_{ \downarrow}, {f^ \dagger}_ \uparrow \rangle \rangle
\end{equation}}}

Solving above coupled equations from equation (\ref{eq:17}) to (\ref{eq:22}) , relevant Grren's function $\langle \langle f_ \uparrow, {f^ \dagger}_ \uparrow \rangle \rangle$ is obtained as follows.
\begin{equation}\label{Greenfn}
   \langle \langle f_ \uparrow, {f^ \dagger}_ \uparrow \rangle \rangle = \frac{1}{\omega-\tilde\epsilon_d-\tilde\Gamma_{2L}-\tilde\Gamma_{2R}-\frac{(\tilde\Gamma_{3L}+\tilde\Gamma_{3R})(\tilde\Gamma_{4L}+\tilde\Gamma_{4R})}{\omega+\tilde\epsilon_d-\tilde\Gamma_{1L}-\tilde\Gamma_{1R}}} 
\end{equation}

where

\begin{equation}
  \tilde\Gamma_{1 \alpha}=\sum_k  |\tilde{h_{k\alpha}}|^2 \left [   \frac{|v_{k\alpha}|^2}{\omega-E_{k\alpha}}+\frac{|u_{k\alpha}|^2}{\omega+E_{k\alpha}} \right ]
\end{equation}

\begin{equation}
  \tilde\Gamma_{2\alpha}=\sum_k  |\tilde{h_{k\alpha}}|^2 \left [   \frac{|u_{k\alpha}|^2}{\omega-E_{k\alpha}}+\frac{|v_{k\alpha}|^2}{\omega+E_{k\alpha}} \right ]
\end{equation}

\begin{equation}
  \tilde\Gamma_{3\alpha}=\sum_k  {\tilde{h_{k\alpha}}}^2 v_{k \alpha}^* u_{k\alpha} \left [   \frac{1}{\omega-E_{k\alpha}}-\frac{1}{\omega+E_{k\alpha}} \right ]
\end{equation}

\begin{equation}
  \tilde\Gamma_{4\alpha}=\sum_k  {\tilde{h_{k\alpha}^*}}^2 u_{k\alpha}^* v_{k\alpha} \left [   \frac{1}{\omega-E_{k\alpha}}-\frac{1}{\omega+E_{k\alpha}} \right ]
\end{equation}

To obtain the parameters b and $\lambda$, we followed the minimization of free energy function, which can be described in terms of density of states [$\tilde\rho_f (\omega)$] as follows.

\begin{equation}
    F=- \frac{1}{\beta} ln Z = - \frac{N}{\beta} \int_{-\infty}^{\infty}  ln(1+e^{-\beta(\omega-\mu)}) \tilde\rho_f(\omega) d \omega
\end{equation}

For a wide flat conduction band of width 2D, change in free energy due to impurity is given as follows ~\cite{Hewson}.
\begin{equation}
    F_{MF} = - \frac{N}{\beta} \int_{-D}^{D}  f(\omega) tan^{-1} (\frac{\tilde \Lambda}{\tilde{\epsilon_d}-\omega}) d \omega + \lambda(b^2-1)
\end{equation}
 where $\tilde\Lambda$ is renormalized hybridization given as follows
 
 \begin{equation}
    \tilde \Lambda = \tilde\Gamma_{2L}+\tilde\Gamma_{2R}+\frac{(\tilde\Gamma_{3L}+\tilde\Gamma_{3R})(\tilde\Gamma_{4L}+\tilde\Gamma_{4R})}{\omega+\tilde\epsilon_d-\tilde\Gamma_{1L}-\tilde\Gamma_{1R}}
\end{equation}
 For flat conduction band, dot-lead coupling becomes k-independent. It is clear that for b=1 and $\lambda=0$ , $\tilde \Lambda$ reduces to that of non interacting hybridization. Minimization of free energy  with respect to parameters b and $\lambda$ , give rise to two coupled integral equations described as follows.
 
\begin{multline} \label{fminb}
    \frac{\partial F_{MF}}{\partial \lambda}=\frac{N}{\pi} \int_{-D}^{D}  \frac{f(\omega)d \omega}{{(\tilde{\epsilon_d}-\omega})^2+{\tilde \Lambda}^2} \left [ \tilde{\Lambda}+ \frac{(\tilde\Gamma_{3L}+\tilde\Gamma_{3R})(\tilde\Gamma_{4L}+\tilde\Gamma_{4R})(\tilde{\epsilon_d}-\omega)}{(\omega+\tilde\epsilon_d-\Gamma_{1L}-\tilde\Gamma_{1R})^2}  \right] \\+ (b^2-1) = 0
\end{multline}
\begin{multline} \label{fminb}
    \frac{\partial F_{MF}}{\partial b}=-\frac{N}{\pi} \int_{-D}^{D}  \frac{f(\omega)d \omega}{{(\tilde{\epsilon_d}-\omega})^2}[(\Gamma_{2L}+\Gamma_{2R}) +\frac{(\Gamma_{3L}+\Gamma_{3R})(\Gamma_{4L}+\Gamma_{4R})}{(\omega+\tilde{\epsilon_d}-\tilde{\Gamma_{1L}}-
    \tilde{\Gamma_{1R}})^2}\\(\lambda+\Gamma_{11L}+\Gamma_{1R}+2b(\omega+\tilde{\epsilon_d}-\tilde{\Gamma_{1L}}-\tilde{\Gamma_{1R}}))]+ (b^2-1) = 0
\end{multline}
In equations (\ref{fminb}) and (\ref{fminlam}), N is level degenracy ( N=2, for single level case) and $f(\omega)$ is distribution function. The Bogoliubonos follow the Fermi-Dirac distribution function for dispersion $E_k$,  given as     follows.
\begin{equation}
    f(\omega)=\langle\beta^\dagger_{k\uparrow}\beta_{k\uparrow}\rangle=\langle\beta^\dagger_{-k\downarrow} \beta_{-k\downarrow}\rangle=\frac{1}{e^{\beta E_k}+1}
\end{equation}

We have used zero temperature for our calculation. To analyse the DOS and energy of ABSs, we need to solve equations (\ref{fminb}) and (\ref{fminlam}) numerically for various parameters of SBMF Hamiltonian.

\section{Results and Discussions}

On the basis of SIAM, we have used Bogoliubov transformation followed by infinite-U slave Boson mean field approach to get the relevant Green's function. From Green's function, density of states of quantum dot is calculated as follows.
\begin{equation}
    \rho_f (\omega)=-\frac{1}{\pi} imag(G_f (\omega^+))
\end{equation}
where, Green's function $G_f (\omega^+)$ is given by equation (\ref{Greenfn}). 

\begin{figure}[ht]
   \begin{minipage}{0.5\textwidth}
     \centering
     \includegraphics[width=1.15\linewidth]{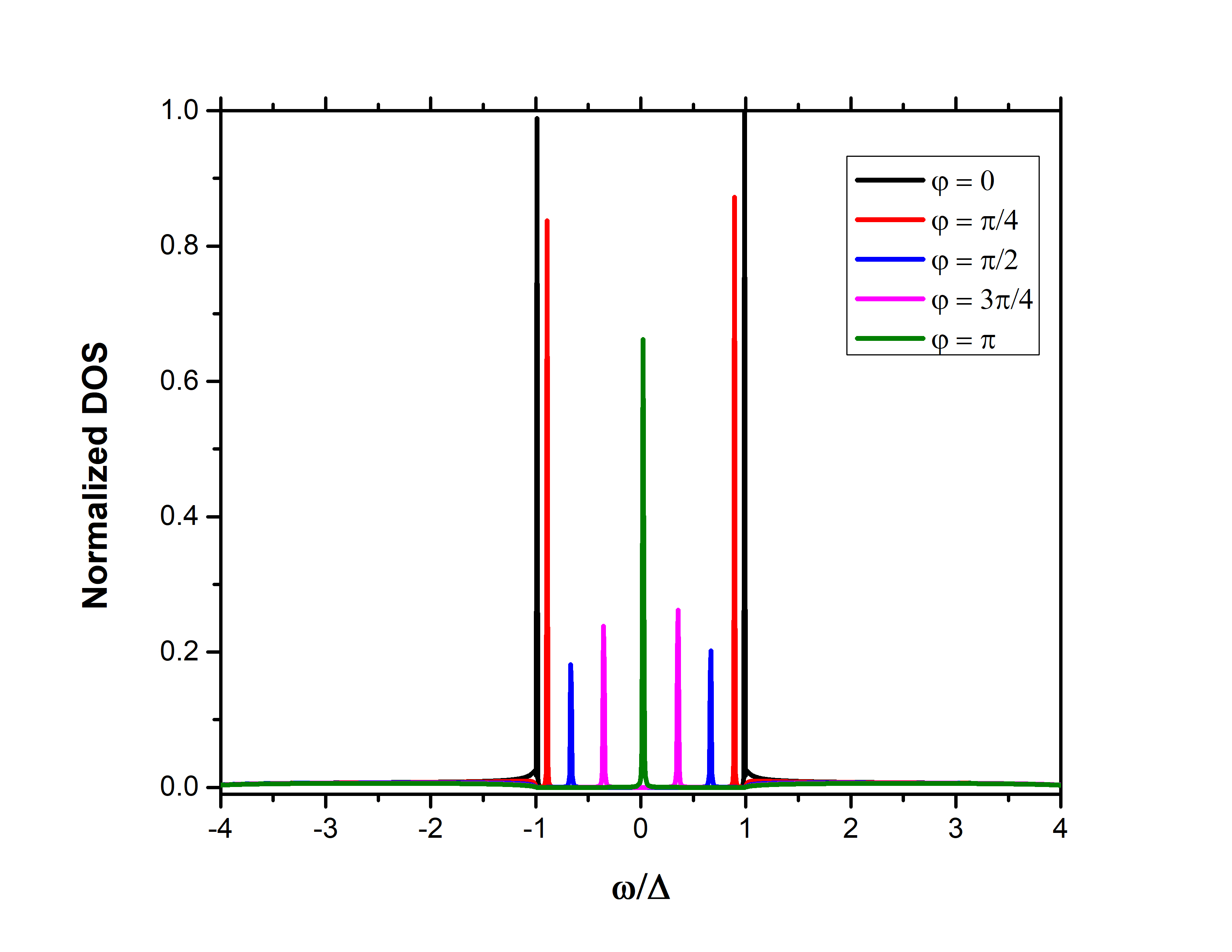} \label{dos_1}
    \subcaption{Normalized density of states in QD}\label{dos_11}
   \end{minipage}\hfill
   \begin{minipage}{0.5\textwidth}
     \centering
     \includegraphics[width=1.15\linewidth]{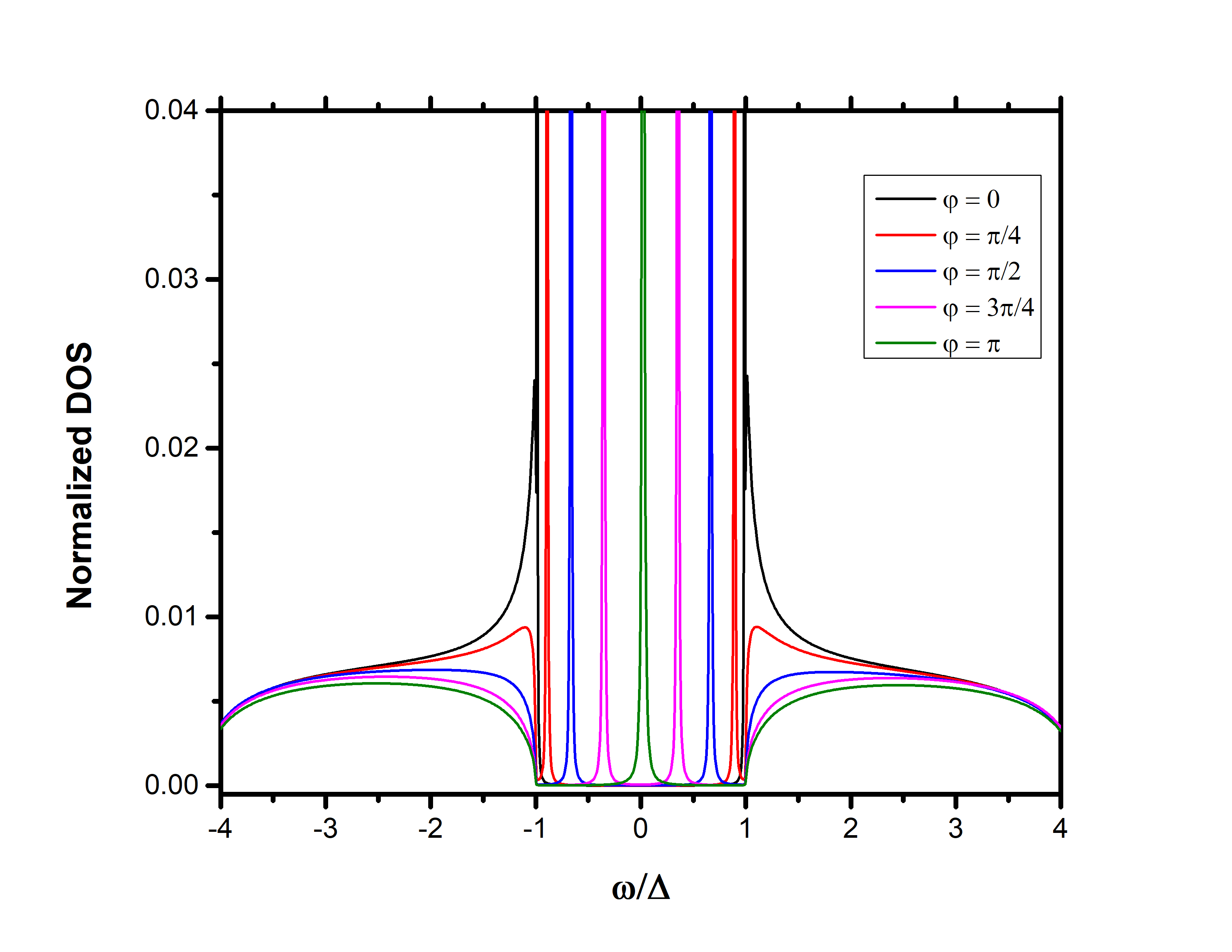} \label{dos_2}
    \subcaption{close up view of normalized DOS in QD}\label{dos_22}
   \end{minipage}
   \caption{Normalized DOS (Density of states) as a function of $\omega/\Delta$ for electron-hole symmetric ($\epsilon_d$ = 0) as well as left-right symmetric ($h_L=h_R$) case  at different values of superconducting phase difference ($\phi=0,\pi/4,\pi/2,3\pi/4,\pi$).}
\end{figure}

In Fig. (\ref{dos_11}), normalized density of states (DOS) is analyzed for identical left and right superconducting lead ($\left|\Delta_L \right|=\left|\Delta_R \right|=1meV$) as a function of $\omega/\Delta$ for various values of superconducting phase difference ($\phi=\phi_L-\phi_R=0,\pi/4,\pi/2,3\pi/4,\pi$) in particle-hole symmetric ($\epsilon_d=0$) and left-right symmetric  ($h_L=h_R$) case. It is observed that in the absence of phase difference between left and right lead ($\phi=0$), superconducting gap can be seen in DOS plot with no subgap states around the Fermi level. As the value of phase difference is increased to $\phi=\pi/4$,  two sub gap states (symmetrically above and below the Fermi level)  appear in the gap, which are called Andreev bound states. Sub gap state below the Fermi level is called lower Andreev bound state and one above the Fermi level is called upper Andreev bound state. In the other words, phase difference introduces paired state  in the quantum dot with reduced gap. Further increasing the value of $\phi$ ($\phi=\pi/2, 3\pi/4$) upper and lower ABSs come closer and closer. For $\phi=\pi$ these two states merge into each other and only a single subgap state is obtained. Fig. (\ref{dos_22}) shows the close up look of Fig. (\ref{dos_11}). Nature of DOS can be observed clearly in Fig. (\ref{dos_22}) at superconducting gap edge. Here, it can be pointed out that the DOS plot for infinite-U case is similar to non interacting case except the amplitude of DOS is decreased. It denotes that transmission of Cooper pair electrons from superconducting lead to QD is decreased. Hence, ABSs are not effected by infinite Coulomb interaction in the dot and still present within the superconducting gap. Dependency of ABSs on phase difference indicates that the transmission of Cooper pairs can be controlled by changing superconducting phase difference between left and right lead. These results are valid in the region $T_K>\Delta$, where $T_K$ is the Kondo temperature corresponding to superconducting lead.

Further, the energy of ABSs is calculated from poles of the Green's function (\ref{Greenfn}), which is given by the solution of non linear equation.

\begin{equation}\label{solabs}
  (\omega_s-\tilde\epsilon_d-\tilde\Gamma_{2L}-\tilde\Gamma_{2R})(\omega_s+\tilde\epsilon_d-\tilde\Gamma_{1L}-\tilde\Gamma_{1R})=(\tilde\Gamma_{3L}+\tilde\Gamma_{3R})(\tilde\Gamma_{4L}+\tilde\Gamma_{4R})  
\end{equation}

\begin{figure}[ht]
   \begin{minipage}{0.5\textwidth}
     \centering
     \includegraphics[width=1.1\linewidth]{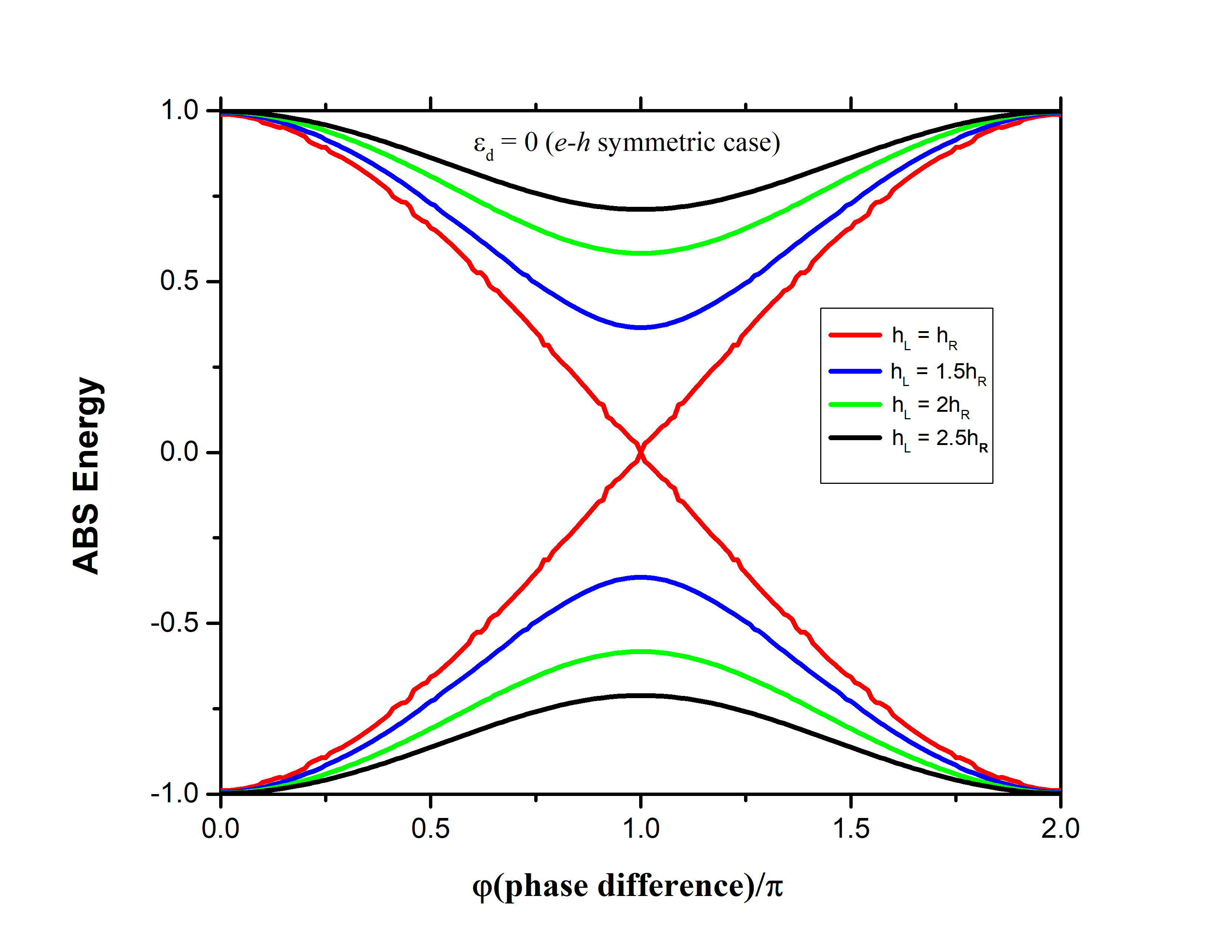} \label{abs1}
    \subcaption{electron-hole symmetric case} \label{abs11}
   \end{minipage}\hfill
   \begin{minipage}{0.5\textwidth}
     \centering
     \includegraphics[width=1.1\linewidth]{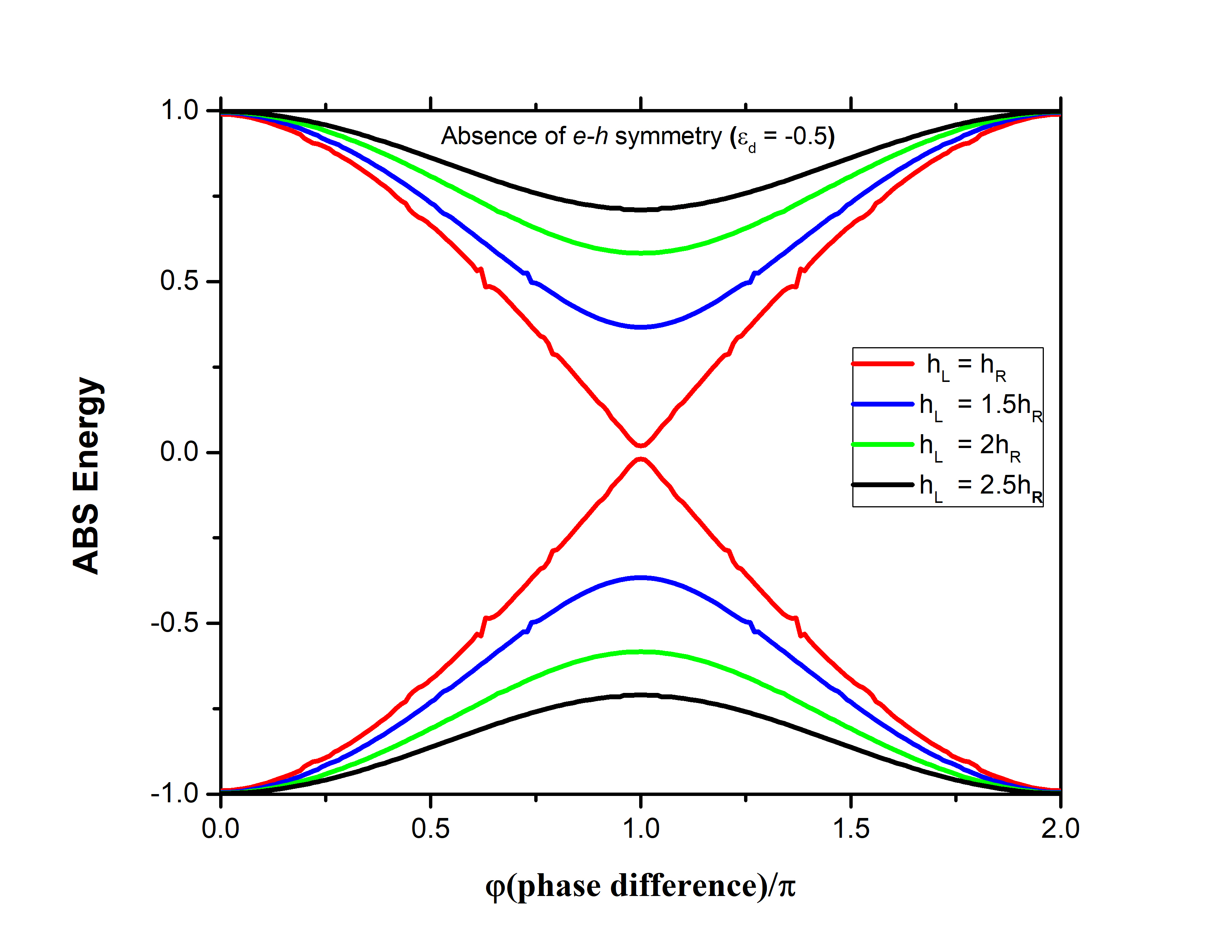} \label{abs2}
    \subcaption{electron-hole non symmetric case} \label{abs22}
   \end{minipage}
   \caption{ABS as a function of phase difference at different dot-lead coupling ratios ($h_L/h_R =1, 1.5, 2, 2.5$).}\label{abs}
\end{figure}

Equation (\ref{solabs}) is solved numerically. In Fig (\ref{abs11}), energy of Andreev bound states is analyzed with variation in superconducting phase difference in electron-hole symmetric case for various left lead-dot coupling ($h_L$) to right lead-dot couling ($h_R$) ratios. For $h_L=h_R$, i.e. left-right symmetry, junction acts as a perfectly transmitting channel. Further increasing the ratio $h_L/h_R$  (1.5,2,2.5) a finite internal gap arises between upper and lower ABSs at $\phi=\pi$. This internal gap increases with increase in the ratio $h_L/h_R$. Fig (\ref{abs22}) shows the energy of ABSs with variation in superconducting phase difference in the absence of electron-hole symmetry for various left lead-dot coupling($h_L$) to right lead-dot coupling ($h_R$) ratios. In this case, there is an small internal gap is present, even at $h_L=h_R$.  Further increasing the ratio $h_L/h_R$  (1.5,2,2.5) width of this finite internal gap is increased. So,it can be pointed out that non-perfect transmitting channel is formed at the junction in the absence of electron-hole symmetry. Due to dependency of ABSs on superconducting phase difference, these are current carrying states. Hence, ABS forms a channel through which Josephson current flows across the S-QD-S junction even at infinite on dot Coulomb interaction limit. These results are in accordance with the works pointed out in other theoretical works, where sub gap states in S-QD-S system is studied ~\cite{ALYeyati}.

\section{Conclusions and outlook}
Employing the infinite-U slave Boson mean field approach for Single impurity Anderson model of superconductor representing S-QD-S junction, sub gap states (Andreev bound states) are observed within the superconducting gap. These sub gap states are found to depend on the superconducting phase difference between two leads as well as left lead-dot coupling to right lead-dot coupling ratio. Due to dependence on phase difference, these states are current carrying states. This formalism can be extended to multilevel quantum dot and also to multi coupled QDs connected to superconducting leads. More, works is required for these types of system because of their potential application in nano electronic devices and hence we are attempting the works in this direction.

 \section*{Aknowledgement}
   One of the authors, Tanuj Chamoli would like to thank Ministry of Human Resource Development (MHRD), Government of India for financial support, received as a research fellowship. 
   \newpage
\bibliography{references}
\bibliographystyle{elsarticle-num}
\end{document}